\documentclass[sigconf]{acmart}

\renewcommand\footnotetextcopyrightpermission[1]{} %
\usepackage{booktabs}
\usepackage{multirow}
\usepackage{pifont}
\usepackage{amsthm}
\usepackage{subcaption}
\newcommand{\code}[1]{{\ttfamily \small #1}}
\usepackage{xcolor}
\definecolor{mygreen}{rgb}{1, 0, 0.6}

\usepackage{hyperref}
\usepackage{amsfonts}
\usepackage{enumitem}
\setlist[itemize]{leftmargin=*}
\setlist[enumerate]{leftmargin=*}

\newcommand{\nBugFixes}{16}
\newcommand{\nParticipants}{27}

\newcommand{\interfaceName}{Code Editing Recorder}
\newcommand{\interfaceNameAbbr}{CER}

\newcommand{\snippet}{buggy code snippet}

\newcommand{\nDebuggingSessions}{98}
\newcommand{\fracPractitioners}{13/27}
\newcommand{\fracAcademics}{11/27}
\newcommand{\fracRemainingPraticipants}{3/27}

\newcommand{\humanMeanAgreement}{0.56}
\newcommand{\recoderMeanAgreement}{0.44}
\newcommand{\sequencerMeanAgreement}{0.35}
\newcommand{\modelvsmodelMeanAgreement}{0.27}

\newcommand{\pvalRecoderVsSequencerTestOnMeanAgreement}{5.08E-6}

\newcommand{\humanAttentionBuggyLinePerc}{36.8\%}
\newcommand{\humanAttentionContextPerc}{63.2\%}
\newcommand{\sequencerAttentionBuggyLinePerc}{67.1\%}

\newcommand{\recoderAttentionBuggyLinePerc}{12.7\%}
\newcommand{\recoderAttentionContextPerc}{87.3\%}

\newcommand{\pvalContextVsCodeLength}{2.30E-4}
\newcommand{\corrContextVsCodeLength}{0.37}

\newcommand{\humanAllPlausiblePatches}{68}
\newcommand{\humanAllPlausiblePatchesPerc}{69.4\%}
\newcommand{\humanAllCorrectPatches}{66}
\newcommand{\humanAllCorrectPatchesPerc}{67.3\%}
\newcommand{\humanAllTotPatches}{98}
\newcommand{\humanAllBugsFixed}{16}
\newcommand{\humanAllBugsFixedPerc}{100.0\%}

\newcommand{\sequencerAllPlausiblePatches}{17}
\newcommand{\sequencerAllPlausiblePatchesPerc}{1.2\%}
\newcommand{\sequencerAllCorrectPatches}{15}
\newcommand{\sequencerAllCorrectPatchesPerc}{1.1\%}
\newcommand{\sequencerAllTotPatches}{1395}
\newcommand{\sequencerAllBugsFixed}{6}
\newcommand{\sequencerAllBugsFixedPerc}{37.5\%}

\newcommand{\sequencerAtFivePlausiblePatches}{2}
\newcommand{\sequencerAtFivePlausiblePatchesPerc}{2.5\%}
\newcommand{\sequencerAtFiveCorrectPatches}{2}
\newcommand{\sequencerAtFiveCorrectPatchesPerc}{2.5\%}
\newcommand{\sequencerAtFiveTotPatches}{80}

\newcommand{\recoderAllPlausiblePatches}{10}
\newcommand{\recoderAllPlausiblePatchesPerc}{1.1\%}
\newcommand{\recoderAllCorrectPatches}{8}
\newcommand{\recoderAllCorrectPatchesPerc}{0.9\%}
\newcommand{\recoderAllTotPatches}{908}
\newcommand{\recoderAllBugsFixed}{5}
\newcommand{\recoderAllBugsFixedPerc}{31.2\%}

\newcommand{\recoderAtFivePlausiblePatches}{2}
\newcommand{\recoderAtFivePlausiblePatchesPerc}{2.5\%}
\newcommand{\recoderAtFiveCorrectPatches}{2}
\newcommand{\recoderAtFiveCorrectPatchesPerc}{2.5\%}
\newcommand{\recoderAtFiveTotPatches}{80}

\newcommand{\meanPercTotAttValidation}{69.3\%}

\newcommand{\nParticipantsValidationExperiment}{ten}
\newcommand{\nValidationSessions}{100}

\theoremstyle{definition}
\newtheorem{definition}{Definition}

\usepackage{tcolorbox}
\newenvironment{answerbox}{
\begin{tcolorbox}[colback=blue!5!white,colframe=blue!5!white,arc=0mm,left=1.5mm,right=1.5mm,top=0mm,bottom=0mm]
}
{
\end{tcolorbox}
}

\AtBeginDocument{%
  }

\acmConference[Arxiv 2023]{Arxiv}{2023}{Stuttgart, DE}

\setcopyright{none}
\begin{document}

\title{Where to Look When Repairing Code? Comparing the Attention of Neural Models and Developers}

\author{Dominik Huber}
\email{dominik@huber-swe.de}
\affiliation{%
  \institution{University of Stuttgart}
  \city{Stuttgart}
  \country{Germany}
}

\author{Matteo Paltenghi}
\email{mattepalte@live.it}
\affiliation{%
  \institution{University of Stuttgart}
  \city{Stuttgart}
  \country{Germany}
}

\author{Michael Pradel}
\email{michael@binaervarianz.de}
\affiliation{%
  \institution{University of Stuttgart}
  \city{Stuttgart}
  \country{Germany}
}

\renewcommand{\shortauthors}{Huber et al.}

\begin{abstract}
  Neural network-based techniques for automated program repair are becoming increasingly effective.
  Despite their success, little is known about why they succeed or fail, and how their way of reasoning about the code to repair compares to human developers.
  This paper presents the first in-depth study comparing human and neural program repair.
  In particular, we investigate what parts of the buggy code humans and two state of the art neural repair models focus on.
  This comparison is enabled by a novel attention-tracking interface for human code editing, based on which we gather a dataset of \nDebuggingSessions{} bug fixing sessions, and on the attention layers of neural repair models.
  Our results show that the attention of the humans and both neural models often overlaps (\sequencerMeanAgreement{} to \recoderMeanAgreement{} correlation).
  At the same time, the agreement between humans and models still leaves room for improvement, as evidenced by the higher human-human correlation of \humanMeanAgreement{}.
  While the two models either focus mostly on the buggy line or on the surrounding context, the developers adopt a hybrid approach that evolves over time, where  \humanAttentionBuggyLinePerc{} of the attention is given to the buggy line and the rest to the context.
  Overall, we find the humans to still be clearly more effective at finding a correct fix, with \humanAllCorrectPatchesPerc{} vs.\ less than 3\% correctly predicted patches.
  The results and data of this study are a first step into a deeper understanding of the internal process of neural program repair, and offer insights inspired by the behavior of human developers on how to further improve neural repair models.
\end{abstract}

\maketitle

\section{Introduction}

To help software developers address the continuously increasing number of bugs, automated program repair (APR)~\cite{cacm2019-program-repair} tries to identify a code transformation that fixes a specific bug without any human intervention.
Over the past few years, a wide range of approaches has been proposed~\cite{yeComprehensiveStudyAutomatic2021}.
An increasing number of them is based on deep neural networks~\cite{vasicNeuralProgramRepair2019, liDLFixContextbasedCode2020b, jiangCURECodeAwareNeural2021, allamanisLearningRepresentPrograms2018a, guptaDeepFixFixingCommon2017}.
Some of these learning-based approaches aim to fix specific bug patterns, such as variable misuse bugs~\cite{dinellaHoppityLearningGraph2019,allamanisLearningRepresentPrograms2018a, vasicNeuralProgramRepair2019}.
Others target a broader range of bugs, such as those contained in QuixBugs~\cite{linQuixBugsMultilingualProgram2017} or Defects4J~\cite{justDefects4JDatabaseExisting2014}, which are popular benchmarks of bugs and corresponding fixes.

Despite the impressive success of recent neural repair models, there still is a very limited understanding of when and why these models work.
We argue that it is crucial not only to investigate their effectiveness on selected benchmarks, but also to study the reasons behind the successes and failures of an approach.
A thorough understanding of the strategies adopted by neural models when reasoning about a repair will be helpful to further improve the current state of the art.
Given that the ultimate goal of APR is to produce bug fixes in a way similar to human developers, a natural question to ask is:
\emph{How do neural APR models compare to humans\\ when both are trying to fix the same bugs?}
Answering this question will help understand how close current APR models are to humans in terms of how to approach the bug fixing problem and in terms of bug fixing effectiveness.
Moreover, better understanding the relationship between human and neural program repair will help design future models that more closely mimic human behavior.

\begin{figure}[t]
  \centering
  \includegraphics[width=\linewidth]{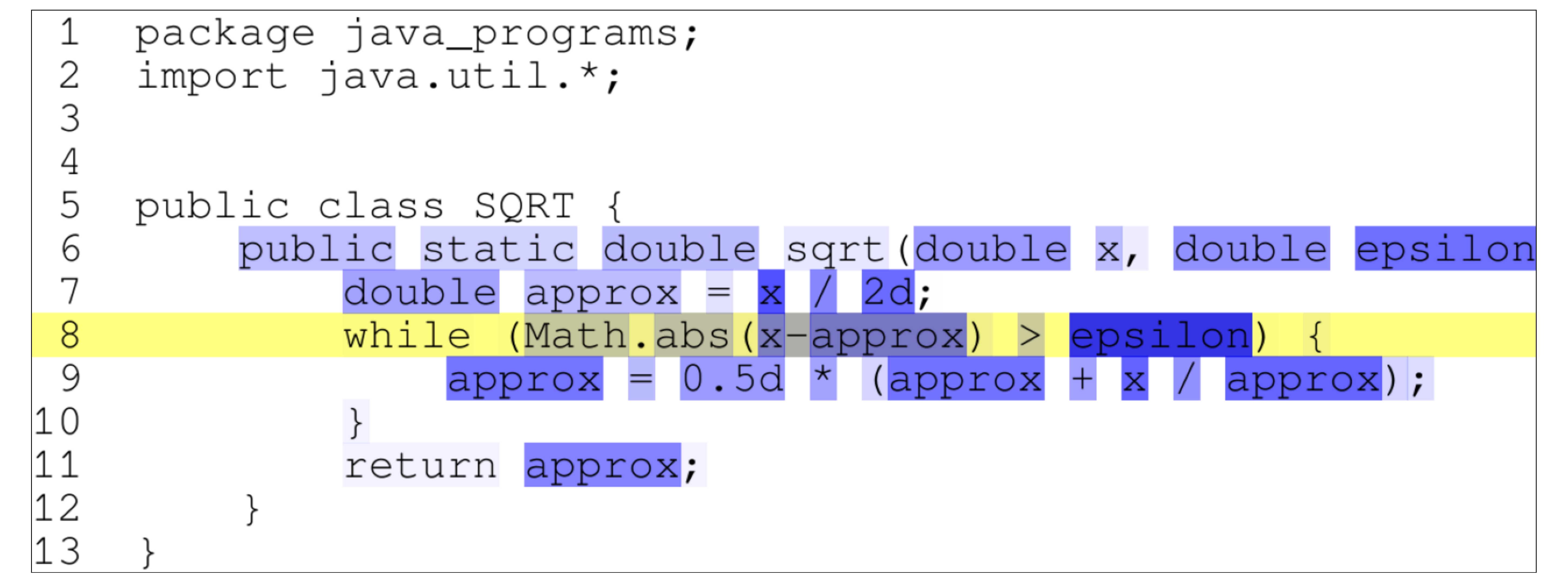}
  \caption{Attention from a neural APR model (Recoder) over a buggy code snippet that computes the square root of a number. The bug is on line 8, where the correct code would be: \code{while (Math.abs(x-approx*approx) > epsilon) {}}.}
  \label{fig:snippet_sqrt}
\end{figure}

This paper addresses the above question through the lens of attention, i.e., what parts of the given code a human or a model focuses on when reasoning about the code.
To capture the attention paid by neural models, we leverage the fact that many recent models implement a special neural network layer devoted to learning where the model should look while making a prediction.
Besides improving the prediction accuracy, this special layer, called \emph{attention layer}~\cite{bahdanauNeuralMachineTranslation2015}, reveals which parts of the code are most relevant for the model to make its predictions and has been widely used to understand how a model makes decisions~\cite{wang2016attention,lee2017interactive,lin2017structured,ghaeiniInterpretingRecurrentAttentionBased2018,ding2017visualizing}.
To capture the attention of human developers, we present a novel programming environment that tracks a developer's attention while repairing a given piece of buggy code.
The interface, called \interfaceName{} (\interfaceNameAbbr{}), blurs most of the code except those parts that are explicitly unblurred by moving the cursor or the mouse.
By tracking blurring and unblurring events we hence can measure which parts of the code a developer focuses on.
To validate the methodology, we also design a novel validation task to measure how effective the \interfaceName{} is at capturing the attention of human developers.

Figure~\ref{fig:snippet_sqrt} shows an example of a buggy program from the QuixBugs dataset~\cite{linQuixBugsMultilingualProgram2017}, with a bug on line~8.
The shading of code tokens shows the attention heatmap of a neural APR model (Recoder~\cite{zhuSyntaxguidedEditDecoder2021}), i.e., which parts of the code the model pays attention to while predicting the fix.
The darker a token, the more attention is spent on it by the model.
In the example, the bug consists in a wrong expression inside the \code{Math.abs()} function, with the correct expression being: \code{x-approx*approx}.

Our study is based on \nDebuggingSessions{} human bug fixing sessions, recorded when fixing \nBugFixes{} bugs from the QuixBugs benchmark~\cite{linQuixBugsMultilingualProgram2017}.
By comparing the data collected from \nParticipants{} human developers to how two neural models, SequenceR~\cite{chenSequenceRSequencetoSequenceLearning2021} and Recoder~\cite{zhuSyntaxguidedEditDecoder2021}, fix the same bugs, we answer the following research questions:
\begin{itemize}
  \item \emph{RQ1: How effective is neural APR compared to human developers?}
  This question is relevant to shed light on the maturity of the neural approaches in relation to their natural competitors.
  We find that the developers clearly outperform both studied neural models.
  For example, the developers come up with a correct fix in
  \humanAllCorrectPatchesPerc{} of their attempts, whereas less than 3\% of the model's predictions yield correct fixes.

  \item \emph{RQ2: To what extent does the attention of developers correlate with the built-in attention layers of neural APR models?}
  This question is relevant for understanding whether humans and models follow a similar reasoning process, and might hint at ways to bring the neural models closer to the humans.
  We find that the attention of both studied models positively correlates with that of the humans, with \sequencerMeanAgreement{} and \recoderMeanAgreement{} correlation for SequenceR and Recoder, respectively.
  At the same time, the agreement between humans and models still leaves room for improvement, as evidenced by the higher human-human correlation of \humanMeanAgreement{}.

  \item \emph{RQ3: How much attention do humans and models pay to the buggy line as opposed to the surrounding code?}
  Addressing this question is relevant as it may influence the design of future neural models and help decide how much context to provide to a model.
  We find that the two studied models follow two very different strategies, with SequenceR paying most attention to the buggy line (\sequencerAttentionBuggyLinePerc{}), whereas Recoder focuses mostly on the context (\recoderAttentionContextPerc{}).
  The human developers strike a balance between both strategies, with \humanAttentionBuggyLinePerc{} focus on the buggy line.

  \item \emph{RQ4: Which types of tokens receive most attention when fixing  a bug?}
  Answering this question could reveal types of tokens that are deemed important by the developers to fix a bug, but that current models disregard.
  We find that the types of tokens that the developers and the models focus on differ significantly.
  For example, human developers focus a lot on operators, while neither of the studied models pays much attention to them.

  \item \emph{RQ5: How does the behavior of the developers evolve during the bug fixing process in terms of focus and code edits?}
  Answering this question will offer a deeper understanding of the relative importance of code comprehension as opposed to code editing.
  We find that the first three quarters of an average human bug fixing session are mostly devoted to understanding the code, whereas the final quarter is dedicated to fixing the code.
  Future repair models could mimic this behavior through neural architectures that explicitly distinguish between the two phases.

\end{itemize}

To the best of our knowledge, this work is the first to compare human developers and neural APR models in a comprehensive study.
Moreover, we are not aware of prior work that systematically studies the effectiveness of developers while fixing bugs in a popular APR benchmark.
A related piece of work compares human and model attention for code summarization~\cite{paltenghiThinkingDeveloperComparing2021}.
This paper differs by considering a different task, different neural models, and by presenting a novel attention-tracking interface specifically designed for code editing.
Our work also relates to prior studies on how developers understand and fix bugs~\cite{castelhanoRoleInsulaIntuitive2019,bohmeWhereBugHow2017a}.
In contrast to our work, these studies do neither consider attention nor compare developers with neural models.

In summary, the contributions of this work are:
\begin{itemize}
  \item An interface for recording human attention and edits performed while fixing a bug, along with evidence that it accurately captures human attention.
  \item A dataset of \nDebuggingSessions{} human bug fixing sessions, including token-level attention maps of developers while fixing bugs in a popular APR benchmark.
  \item A systematic comparison of human developers and neural APR models in terms of their overall effectiveness and the attention spent during the repair process.
  \item Insights into commonalities and differences between human developers and neural APR models, which could help design future neural models that even more closely mimic the human repair process.
\end{itemize}

We envision our results to serve as a basis for further studying the strengths and weaknesses of neural APR models, and to provide insights that guide the development of more accurate and explainable future models.

\section{Background}

\subsection{Automated Program Repair}

Automated program repair (APR) means techniques designed to fix bugs with no or little human intervention.
Current techniques can be broadly categorized into three kinds of approaches~\cite{cacm2019-program-repair}: (1) heuristic repair, such as GenProg~\cite{legouesGenProgGenericMethod2012}, which generates many patches via genetic programming and validates them via a test suite, (2) constraint-based repair, such as SemFix~\cite{Nguyen2013b}, which extracts constraints from the existing buggy code and generates a new patch based on them, (3) learning-based repair, such as DeepFix~\cite{guptaDeepFixFixingCommon2017}, where a neural model of code is trained with a large dataset of bug-fixes in an end-to-end fashion.
This paper studies neural models for learning-based repair.

Orthogonal to the kind of approach is the question what information is provided to an APR technique in addition to the buggy program.
Some approaches assume to have only the buggy code, such as Hoppity~\cite{dinellaHoppityLearningGraph2019}, a learning-based approach that learns to localize the buggy line and then propose a fix.
Others, such as Getafix~\cite{oopsla2019}, assume to also know the location of the bug.
The perhaps largest group of approaches, such as GenProg~\cite{legouesGenProgGenericMethod2012}, assumes to know not only the location of the bug, but to also have access to a test suite to validate a possibly large number of candidate fixes until finding one that passes all tests.
This paper studies APR techniques that assume to know the bug location and rely on test suite-based validation.

Another dimension to classify APR approaches is the class of bugs they address.
Some approaches specialize on few specific kinds of bugs, such as variable misuse~\cite{allamanisLearningRepresentPrograms2018a}, while others target a broader range of bugs.
This paper studies APR techniques that focus on a wide range of bugs, as long as they are fixable by editing a single line of code.

\subsection{Attention Mechanism}

The attention mechanism is a type of neural network layer that indicates how relevant a specific part of the input is for making a prediction.
Introduced by Bahdanau et al.~\cite{bahdanauNeuralMachineTranslation2015}, attention now is part of many modern neural network architectures, including those used for neural APR.
Attention can be represented as a function $f_{\mathit{att}}$ from a sequence of $n$ inputs, each represented with a $d$-dimensional embedding vector, to a vector of attention weights:
$
  f_{\mathit{att}}: \mathbb{R}^{d \times n} \rightarrow \mathbb{R}^n
$.
For models of code, the sequence of inputs represents, e.g., programming language tokens or a serialization of nodes in an abstract syntax tree.

Three forms of attention are relevant in this study.
First, \textit{global attention} allows a model to explicitly focus on some part of the input when predicting the output, which is commonly used in encoder-decoder architectures, such as the sequence-to-sequence model of SequenceR~\cite{chenSequenceRSequencetoSequenceLearning2021}.
Second, \textit{self-attention} is the fundamental building block of transformers~\cite{vaswaniAttentionAllYou2017}, where attention is used also in the encoder.
Specifically, each element in the input sequence of length~$n$ gets encoded as the weighted sum of its own embedding and the embeddings of other elements, selected based on attention weights, in the sequence.
To compute the overall attention paid to the input sequence, the element-level attention weights can be combined to produce a single vector of length $n$.
Self-attention is used, e.g., in the Recoder APR model~\cite{zhuSyntaxguidedEditDecoder2021}.
Third, \textit{copy attention} is used by some models, e.g., SequenceR~\cite{chenSequenceRSequencetoSequenceLearning2021}, to select which tokens to copy verbatim from the input to the output,which is helpful to tackle the vocabulary problem~\cite{karampatsisBigCodeBig2020}.

Beyond improving model accuracy, the attention mechanism offers a view into otherwise mostly black-box neural models, by showing which parts of an input a model considers to be most relevant~\cite{wang2016attention,lee2017interactive,lin2017structured,ghaeini2018interpreting,ding2017visualizing}.
Our study uses this property of attention to compare neural models with human developers.

\section{Methodology}

We start defining the program repair task (Section~\ref{sec:task_design}), how we capture the attention and edits of human developers (Section~\ref{sec:methodology human repair}), and how we extract the attention of neural APR models (Section~\ref{sec:neural_models_selection}).

\subsection{Program Repair Task}
\label{sec:task_design}
\label{sec:dataset_selection}

The task to be addressed by human developers and neural models is to fix bugs in a given piece of code.
In line with the focus of many neural APR techniques, and to keep the task manageable for the developers, we focus on bugs that can be fixed by editing a single line.
As a dataset of bugs, we pick QuixBugs~\cite{linQuixBugsMultilingualProgram2017} because it offers self-contained problems that, in contrast to, e.g., the Defect4J~\cite{justDefects4JDatabaseExisting2014} bugs, do not require prior knowledge of a complex project.
The bugs are in Java code, which comes with test cases that we use to validate whether a fix is plausible.
Since the study involves a significant time investment from our human participants, we select a subset of \nBugFixes{} of the QuixBugs bugs for the whole experiment.

\subsection{Human Program Repair}
\label{sec:methodology human repair}

\subsubsection{\interfaceName{} (\interfaceNameAbbr{})}
\label{sec:interface}

\begin{figure}[t]
  \centering
  \includegraphics[width=\linewidth]{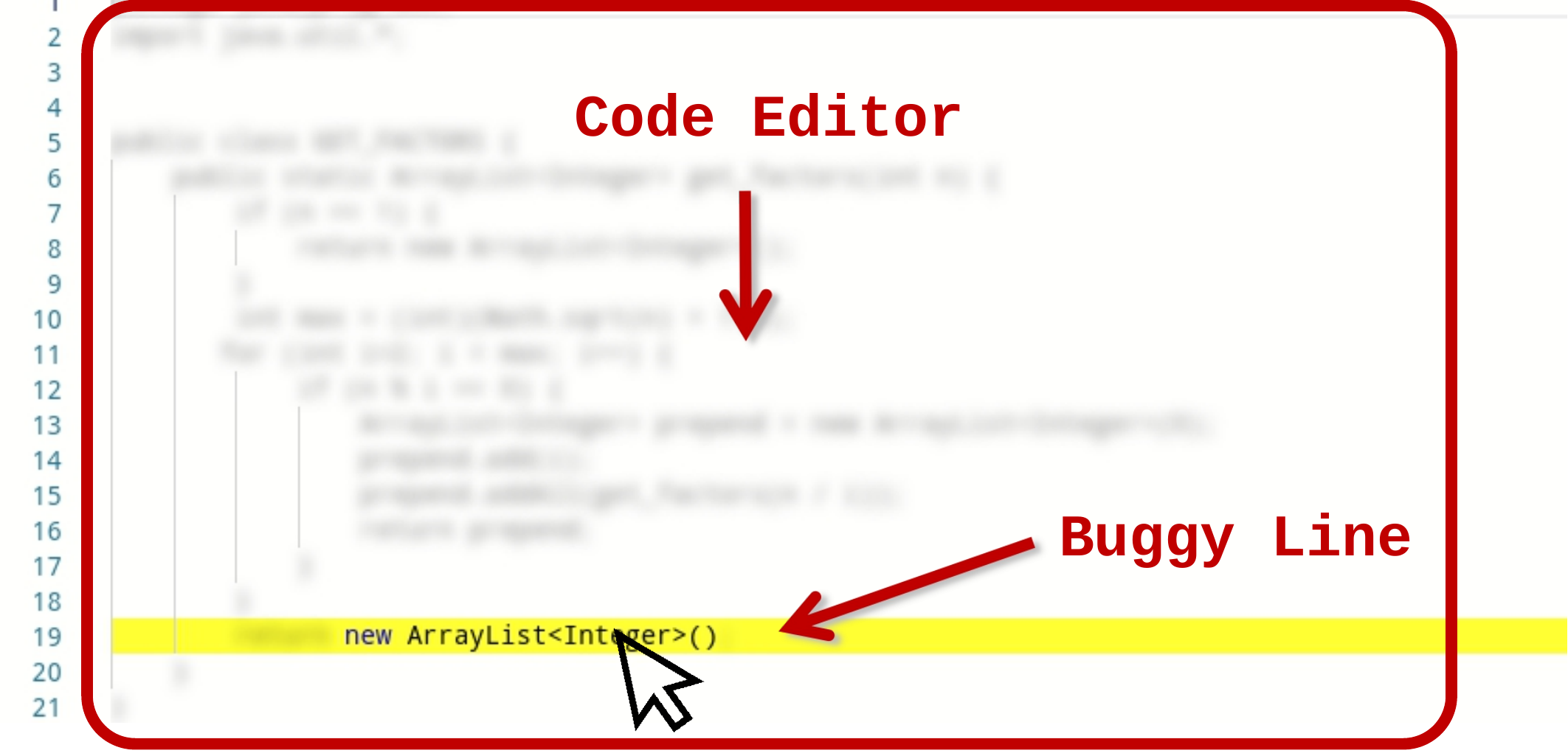}
  \caption{Code editor area of the \interfaceNameAbbr{}.}
  \label{fig:interface}
\end{figure}

To capture how human developers address the program repair task, we present a novel programming interface built specifically for this purpose.
Figure~\ref{fig:interface} shows the code editing area of the interface, which contains the source code to be fixed.
We build upon the browser-based Monaco editor\footnote{\url{https://microsoft.github.io/monaco-editor/}} used by VSCode\footnote{\url{https://code.visualstudio.com/}}, providing an editing experience similar to what developers might usually use.
As comparing different APR approaches is difficult when they make different assumptions about whether the bug location is known~\cite{liuYouCannotFix2019}, we assume perfect fault localization and focus on the patch generation step.
Hence, the buggy line is shown to the developers by highlighting it in yellow.
To prevent developers from editing any other lines, \interfaceNameAbbr{} restricts editing to the buggy line only.
Despite this restriction on editing, the developer can, of course, freely explore all parts of the code.
In addition to the editor, \interfaceNameAbbr{} shows information about the algorithm that the code intends to implement, and a button to submit the solution or indicate that the developer does not know how to fix the bug.

\subsubsection{Capturing Human Attention}
\label{sec:attention_study}
\label{sec:dev_attention_extraction}

To track the attention of developers while they inspect and edit the code, we instrument the code editor using an unblurring mechanism.
Initially, the editor blurs all code tokens and temporarily unblurs only those tokens onto which the developer moves either the keyboard cursor or the mouse, and those in their vicinity.
Similar to prior work, which uses unblurring to capture human attention during code summarization~\cite{paltenghiThinkingDeveloperComparing2021}, we set the size of the unblurred neighborhood to three tokens before and three tokens after the cursor, and the interface never unblurs more than one line at the time.
In contrast to any prior work, \interfaceNameAbbr{} goes beyond a pure code inspection task, and also tracks the developers while they are editing the code.
In particular, our interface shows not just a static image of the code, but combines unblurring-based attention tracking with an interactive code editor.

\interfaceNameAbbr{} records the time spent by the developer in the code editor and captures the following information:

\begin{definition}[Developer attention record]
	The developer attention record is a tuple
	$
	  \mathit{dr} = (\mathit{id}, \mathit{uid}, \mathit{events}, l)
	$,
	where $\mathit{id}$ is a unique identifier for the code to fix,
	$\mathit{uid}$ is a unique identifier for the developer,
	$\mathit{events}$ is a sequence of interaction events,
	and $l$ is the ``fix done'' or ``cannot fix'' label of the submission.
\end{definition}

We record three types of interaction events.
The first is the \textit{unblur event}, which is triggered when the developer moves the cursor or the mouse on some tokens to make them visible.
This event contains information on the currently visible tokens.
The second is the \textit{blur-everything event}, which gets triggered after three seconds of inactivity and signals the moment when all tokens get blurred again.
Finally, the \textit{edit event} is recorded every time that the buggy line is edited.

An interesting challenge in the editing recording arises because the code edits could insert, change or remove tokens.
This sometimes creates ambiguity, such as unparsable code due to unfinished code tokens, or multiple tokens with the same content, making it difficult to understand which token was present in the original code and which token has been newly created.
To handle this ambiguity, \interfaceNameAbbr{} tracks the time a token is visible only for tokens in the original code, and does not track attention on updated or newly inserted tokens.
This way to capturing attention is consistent with the way APR models look at the code, since they typically distribute their attention over the original code rather than over the edited code.

Given the events recorded by the \interfaceNameAbbr{} interface, we convert them to an attention vector, which assigns to each token a value indicating how much attention is paid to that token by the developer.
We approximate the attention given to a token as the total time a token is visible, analogous to the fixation time used in eye tracking~\cite{guarneraEnhancingEyeTracking2019, sharafiPracticalGuideConducting2020} and to the human attention tracking proposed by Paltenghi et al.~\cite{paltenghiThinkingDeveloperComparing2021}:

\begin{definition}[Developer attention]
Suppose a developer attention record $
  \mathit{dr} = (\mathit{id}, \mathit{uid}, \mathit{events}, l)
$
for a buggy code snippet with $n$ tokens.
The developer attention vector is $d = (d_1, d_2, ..., d_n)$, where $d_i$
is the time, normalized to the total time of the record, that the token at position $i$ has been visible to the participant according to the interaction events.
\end{definition}

An alternative to capturing human attention via the \interfaceNameAbbr{} interface is eye tracking~\cite{castelhanoRoleInsulaIntuitive2019,Rodeghero2015,Fakhoury2021}.
Both methodologies assume that a participant's attention can be captured by where the participant looks.
The main benefit of an unblurring-based interface, such as \interfaceNameAbbr{}, is its ease of deployment, which does not require any special equipment and even allows for remote participation.
Experiments that compare eye tracking with an unblurring-based interface similar to \interfaceNameAbbr{} show that both provide similar results~\cite{DBLP:journals/tochi/KimBBGODP17}.

\subsubsection{Participants}
\label{sec:recruiting}

As participants of our study, we recruit \nParticipants{} developers.
All participants are recruited via personal contacts to reduce the risks of involving unqualified  or unmotivated participants, which may easily happen with crowdsourcing platforms.
The participants are composed of both software professionals and academics, forming respectively \fracPractitioners{} and \fracAcademics{} of all participants.
The remaining \fracRemainingPraticipants{} have decided to not share their background.
In terms of the highest level of education, 18/27 had a B.Sc. degree and 8/27 a M.Sc. degree.
Moreover, 16/27 had five or more years of programming experience, 9/27 had between two and four years of programming experience.

Similarly to previous work~\cite{paltenghiThinkingDeveloperComparing2021,wainakhIdBenchEvaluatingSemantic2021}, we aim at collecting at least five human bug-fixes for each \snippet{}.
For the orthogonal decision on how many bugs a single annotator should inspect, we run a pilot study and benchmarked the time required to fix a QuixBugs-like bug using \interfaceNameAbbr{}.
The pilot study showed that 15 minutes are usually enough to address a single bug.
We thus assign to each participant four bugs, giving a total time of one hour for participating in the study.

\subsection{Neural Program Repair}
\label{sec:neural_models_selection}

\subsubsection{Models}

For this study, we focus on two neural models that fulfill the following selection criteria: (1) they use an attention mechanism to make predictions, (2) they have different neural architectures and code representations, and (3) they recently achieved state of the art results, (4) they must be publicly available.
Given these criteria, we select SequenceR~\cite{chenSequenceRSequencetoSequenceLearning2021} and Recoder~\cite{zhuSyntaxguidedEditDecoder2021}.
The two models are representative for different kinds of learning-based repair approaches.
SequenceR uses global attention, whereas Recoder uses self-attention.
They also have different types of architectures, as SequenceR is based on a sequence-to-sequence model with bidirectional LSTMs, whereas Recoder builds on an AST-based transformer~\cite{sunTreeGenTreeBasedTransformer2020}.
Both models are queried in perfect localization mode, thus they know which is the buggy line when generating patches, similarly to the human setup.

\subsubsection{Extracting Attention}
\label{sec:apr_attention_extraction}

For SequenceR~\cite{chenSequenceRSequencetoSequenceLearning2021}, we extract the attention on a token per token basis.
For each code snippet, we let the model predict a fix by generating a sequence of output tokens that greedily picks the next token based on the highest assigned probability.
Then, for each output token, we record the global attention, which yields a vector of length equal to the input code.
We then average all the vectors extracted from all output tokens into a single attention vector.
To ensure that the attention vector from SequenceR matches those collected from the humans, we disable a preprocessing step of SequenceR that applies semantics-preserving code transformations, such as replacing \code{else if () \{...\}} with \code{else \{ if (...) \{...\}\}}.

For Recoder~\cite{zhuSyntaxguidedEditDecoder2021}, we extract an attention vector from the first self-attention transformer layer of the ``code reader'' component.
The reason is that the first layer pays direct and exclusive attention to the input program, avoiding the problem that information from different input elements gets increasingly mixed up in later layers~\cite{abnarQuantifyingAttentionFlow2020}.
Recoder represents the input program as an AST and then reasons about the sequence of its nodes extracted via depth-first traversal (Section~\ref{sec:neural_models_selection}).
Hence, to convert the attention vector on AST nodes into an attention vector on tokens, we map all AST nodes to their corresponding token ranges.
Specifically, each AST node gets mapped to the terminal tokens that are direct or indirect children of that node, including tokens that do not explicitly appear in the AST, such as parenthesis.
Given this mapping, we compute token-level attention by equally distributing the node-level attention across all tokens a node maps to.
For example, an expression \code{a + b} is represented by a \textit{binary expression} node with \code{+} attribute and two children: identifier nodes \code{a} and \code{b}.
Supposing an attention weight of 15 on the binary expression node, each of the three tokens (\code{a}, \code{+}, \code{b}) gets 5 points of attention.
If a terminal node, e.g., \code{a}, also receives some direct attention, then this gets added to the attention propagated from its parent nodes.

Both models generate the patch via a sequential process that produces tokens (for SequenceR) or edit actions (Recoder).
We summarize the attention a model pays when predicting its output into a single attention vector as follows.

\begin{definition}[Neural APR attention]
Suppose a \snippet{} with $n$ tokens and a sequence of $k$ predicted output elements, where the model has an attention vector $a^j$ for each predicted output element.
The neural APR attention vector is $
m = (m_1, m_2, ..., m_n)
$, where $m_i = mean(a^1_i, ..., a^k_i)$, with $a^j_i$ indicating the attention weight assigned by the model to the
token at position $i$ of the \snippet{} during the prediction
of the $j$-th predicted output element.
\end{definition}

In particular, Recoder produces the same attention vector $v$ on the input code for all $n$ predicted output elements of a single patch, thus computing the mean across the $n$ attention vectors returns the same single vector $v$.
Recently proposed alternative methods to post-process attention weights~\cite{abnarQuantifyingAttentionFlow2020,paltenghiExtractingMeaningfulAttention2022} are designed for transformer-based models only, and hence, cannot be used for Recoder.

\section{Validation of Methodology}
\label{sec:methodology_validation}

Our methodology is based on the hypothesis that measuring human attention via \interfaceNameAbbr{} (Section~\ref{sec:dev_attention_extraction}) accurately measures what part of the given code the developers focus on.
The following validates this hypothesis by comparing the measured human attention against an objective reference.
To this end, we design a question-answering task on code where the expected area of interest is known a priori, and check whether \interfaceNameAbbr{} collects attention records with a strong focus on that area.
The requirements for such a task are that: (1) the question cannot be answered without looking at the expected area of interest, (2) it requires continuous focus on the expected area of interest, so that there is a strong attention signal from the human to be captured, and (3) the area of interest can be found quickly, to reduce bias depending on the retrieval abilities of the different participants, (4) the area of interest spans a small number of lines, which reduces the risk of accidentally paying attention to it.

Given these requirements, we design ten tasks that each consist of a Java code snippet and a question.
The code snippets are randomly sampled from five Java projects in Defect4J~\cite{justDefects4JDatabaseExisting2014}: \code{jfreechart}, \code{commons-math}, \code{commons-lang}, \code{commons-collections}, and \code{commons-io}.
The length of the snippets ranges from 23 to 60 lines of code, with an average of 47.1 lines, thus roughly similar to the length of snippets in the main experiment, which range from 12 to 56 lines with an average of 24 lines.
Two examples of questions are: (1) \textit{How many brackets are used in the regex GSON\_VERSION\_PATTERN, where (, ), [, ], \{, \}  are all brackets.}; (2) \textit{In the test method testBadAdditive(), there is a function call with some arguments.
Report the list of all numeric arguments in decreasing order. (E.g., [3,2,1].)}.
Both questions require the user to quickly spot the relevant area of interest and focus on it for a while.
We ask \nParticipantsValidationExperiment{} participants to each answer the ten questions, which yields \nValidationSessions{} attention records.

\begin{figure}[t]
  \centering
  \includegraphics[width=\linewidth]{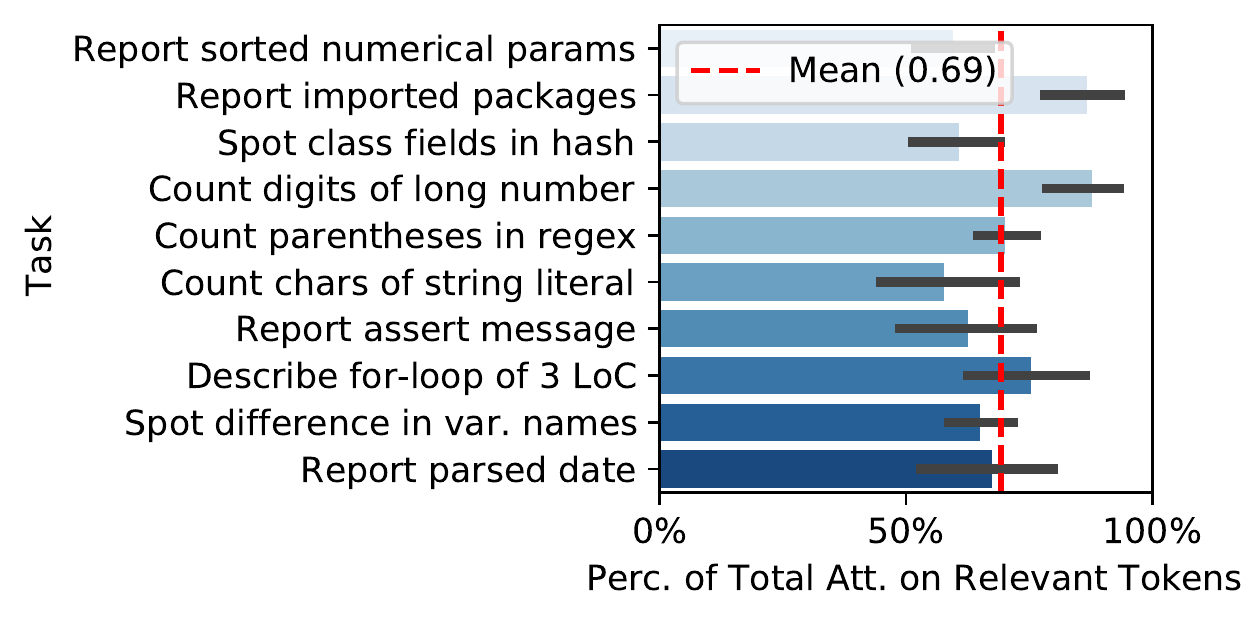}
  \caption{Percentage of total attention given to the relevant area of interest.}
  \label{fig:validation_perc_tot_att}
\end{figure}

We compare the \nValidationSessions{} attention records to the known relevant area of interest.
Figure~\ref{fig:validation_perc_tot_att} shows the percentage of total attention given to the relevant area of interest for each participant.
The results show that, despite the area of interest having a limited number of lines of code (median: 2), \interfaceNameAbbr{} is able to capture a strong attention signal from the developers, with an average of \meanPercTotAttValidation{} of the total attention given to the relevant area of interest.
We conclude from this experiment that our methodology captures human attention with high accuracy, providing a solid basis for our study.

\section{Results}

\subsection{RQ1: Effectiveness of Developers and Neural Models}

Newly proposed neural APR approaches are usually evaluated against existing automated approaches.
Instead, RQ1 investigates how neural approaches compare to developers in terms of their effectiveness at finding a suitable fix.
We answer this question using three metrics:
(1) \textit{Absolute fixing accuracy}: For how many bugs does an ``approach'', i.e., either the set of developers or a model, suggest at least one correct fix?
(2) \textit{Plausible fix ratio}: How many of all proposed patches pass the tests suite?
(3) \textit{Correct fix ratio}: How many of all proposed patches are correct fixes?
For (1) and (3), the term ``correct'' refers to fixes validated by first running the test suite and by then manually checking that a fix indeed matches the expected behavior.

\begin{table}[t]
  \caption{Individual bugs and their absolute fixing accuracy. \ding{51} and - mean that a correct fix is found and not found, respectively. Legend: WEO (Wrong expression operator), IV (Incorrect variable), MC (Missing condition), ICO (Incorrect comparison operator), MA+1 (Missing/added +1), MFC (Missing function call), MAE (Missing arithmetic expression).}
  \label{tab:bug_list}
  \begin{tabular}{@{}lp{0.05\textwidth}p{0.07\textwidth}p{0.06\textwidth}@{}}
    \toprule
    Algorithm (Type of bug) & Devs & SequenceR & Recoder \\
    \midrule
                        Bitcount (WEO) &      \ding{51} &     - &   \ding{51} \\
                     Bucket sort (IV) &      \ding{51} &     \ding{51} &   - \\
     Linked list cycle detection (MC) &      \ding{51} &     - &   - \\
             Fancy binary search (ICO) &      \ding{51} &     \ding{51} &   \ding{51} \\
                   Binary search (MA+1) &      \ding{51} &     - &   - \\
             Prime factorization (MFC) &      \ding{51} &     - &   - \\
                 Towers of Hanoi (IV) &      \ding{51} &     \ding{51} &   \ding{51} \\
                      K-heap sort (MFC) &      \ding{51} &     - &   - \\
                     QuickSelect (MAE) &      \ding{51} &     - &   - \\
            Levenshtein distance (MA+1) &      \ding{51} &     \ding{51} &   \ding{51} \\
      Longest common subseq. (MFC) &      \ding{51} &     - &   - \\
                 Maximum sublist sum (MFC) &      \ding{51} &     - &   - \\
                      Merge sort (MA+1) &      \ding{51} &     \ding{51} &   - \\
                Next permutation (ICO) &      \ding{51} &     \ding{51} &   \ding{51} \\
                   Making change (MC) &      \ding{51} &     - &   - \\
                     Square root (MAE) &      \ding{51} &     - &   - \\
    \midrule
    Absolute fixing accuracy &
    \humanAllBugsFixed{}/16&
    \sequencerAllBugsFixed{}/16&
    \recoderAllBugsFixed{}/16\\
     & (\humanAllBugsFixedPerc{}) & (\sequencerAllBugsFixedPerc{}) & (\recoderAllBugsFixedPerc{}) \\

    \bottomrule
    \end{tabular}
\end{table}

Table~\ref{tab:bug_list} shows the list of \snippet{}s under study, together with the type of bug, and whether the developers and the neural APR models are able to find a fix for the bug.
In the last row, we report the absolute fixing accuracy for developers, SequenceR, and Recoder.
The developers are clearly more effective, as at least one of them finds a correct fix for each of the bugs.
In contrast, SequenceR and Recoder fix six and five of the 16 bugs, respectively.\footnote{Our results for both SequenceR and Recoder slightly differ from those reported in the respective papers. For Sequencer, the reason is the code preprocessing we remove to compute token-level attention (Section~\ref{sec:apr_attention_extraction}). For Recoder, we attribute the differences to non-determinism in the neural model.}
Note that the goal of our study is not to compare different APR techniques with each other.
Instead, we refer to Zhu et al.~\cite{zhuSyntaxguidedEditDecoder2021} for a detailed comparison between Recoder and SequenceR.

\begin{table}
  \caption{Comparison of plausible and correct patches.}
  \label{tab:performance}
  \setlength{\tabcolsep}{3pt}
  \begin{tabular}{@{}lrr|rr@{}}
    \toprule
    &\multicolumn{2}{c}{Plausible patch ratio} & \multicolumn{2}{c}{Correct patch ratio}\\
    \midrule
    \midrule
    & Top-5 & Top-100 & Top-5 & Top-100 \\
    \midrule
    SequenceR & \sequencerAtFivePlausiblePatches{}/\sequencerAtFiveTotPatches{}~(\sequencerAtFivePlausiblePatchesPerc{})& \sequencerAllPlausiblePatches{}/\sequencerAllTotPatches{}~(\sequencerAllPlausiblePatchesPerc{})& \sequencerAtFiveCorrectPatches{}/\sequencerAtFiveTotPatches{}~(\sequencerAtFiveCorrectPatchesPerc{})& \sequencerAllCorrectPatches{}/\sequencerAllTotPatches{}~(\sequencerAllCorrectPatchesPerc{})\\
    Recoder & \recoderAtFivePlausiblePatches{}/\recoderAtFiveTotPatches{}~(\recoderAtFivePlausiblePatchesPerc{})& \recoderAllPlausiblePatches{}/\recoderAllTotPatches{}~(\recoderAllPlausiblePatchesPerc{})& \recoderAtFiveCorrectPatches{}/\recoderAtFiveTotPatches{}~(\recoderAtFiveCorrectPatchesPerc{})& \recoderAllCorrectPatches{}/\recoderAllTotPatches{}~(\recoderAllCorrectPatchesPerc{}) \\
    \midrule
    \midrule
    & \multicolumn{2}{c|}{5-7 developers/bug} & \multicolumn{2}{c}{5-7 developers/bug} \\
    \midrule
    Developers & \multicolumn{2}{c|}{\humanAllPlausiblePatches{}/\humanAllTotPatches{}~(\humanAllPlausiblePatchesPerc{})} & \multicolumn{2}{c}{\humanAllCorrectPatches{}/\humanAllTotPatches{}~(\humanAllCorrectPatchesPerc{})} \\
  \bottomrule
\end{tabular}
\end{table}

Table~\ref{tab:performance} reports the plausible fix ratio and the correct fix ratio for the two models (top) and the developers (bottom).
Both models predict a ranked list of $k$ fix candidates, which are assumed to be validated against the test suite to decide which fix to suggest to a user.
We evaluate two values of $k$ for each model, top-5 and top-100, and measure for all candidates predicted within the top-$k$ how many of them are plausible or correct.
For example, when considering the top-5, SequenceR predicts 16 bugs $\times$ 5 = 80 different fix candidates.
Out of these 80 fix candidates, two are correct.
For top-100, the number of predicted fix candidates is below $100\times16=1,600$ because of predictions that contain the ``unknown'' token (for SequenceR) and because of duplicate predictions (for Recoder).
For the developers, we check all fixes the developers propose during the \nDebuggingSessions{} bug fixing sessions, which means between five and seven fix candidates per bug.
Filtering the developer-provided fixes to remove invalid session where the participants admitted using an external source (e.g., StackOverflow or GitHub Copilot) to find a fix, reduces them from 4$\times \nParticipants=108$ to \nDebuggingSessions{}.

The key take-away of Table~\ref{tab:performance} is that the majority of the developer-proposed fixes are plausible (\humanAllPlausiblePatchesPerc{}) and most of them also correct (\humanAllCorrectPatchesPerc{}).
In contrast, the large majority of the model-predicted fix candidates are neither plausible nor correct, with success ratios of less than 3\%.

\begin{answerbox}
\textbf{Answer to RQ1}: Human developers are clearly more effective than the studied neural APRs models, confirming that, when fixing the \snippet{}s of QuixBugs, these models have not yet reached developer-level effectiveness.
\end{answerbox}

\textit{Implications:}
Our results show that state of the art neural APR models are still far from being a full replacement for developers, even in a relatively well controlled environment and on single-line bugs.
These results motivate further work to improve future neural APR models.

\subsection{RQ2: Correlation of Human and Model Attention}
\label{sec:correlation results}

Observing and comparing the attention patterns of developers and neural APRs models might reveal interesting differences in their bug fixes strategies.
To this end, we investigate the correlation between the attention vectors of developers and models as explained in Section~\ref{sec:attention_study}.

We quantify correlation using the Spearman rank correlation coefficient~\cite{spearmanProofMeasurementAssociation1904}.
This coefficient measures the strength and direction of association between two ranked lists, which in this case are the tokens in the code sorted by their attention:
$\mathit{Spearman} = \frac{cov(rg_1, rg_2)}{\sigma_{rg_1},\sigma_{rg_2}}$ with $rg_1$ and $rg_2$ being the ranks of the two attention vectors, e.g., observed from a developer and a model.
Being a statistical test, the metric provides a p-value, which gives a convenient way to filter out data where the correlation is not statistically significant.
We adopt a significance threshold of 0.05 and discard all data points with a higher p-value.
Inspired by related work on attention~\cite{sharmaExploratoryStudyCode2022,Jain2019}, we also compute the Jensen-Shannon divergence~\cite{journals/tit/Lin91} to measure the distance between the attention vectors.
The results are consistent with those obtained via Spearman, but lack a p-value, and we hence focus on the Spearman rank correlation coefficient in the following.

\begin{figure}[t]
  \centering
  \includegraphics[width=\linewidth]{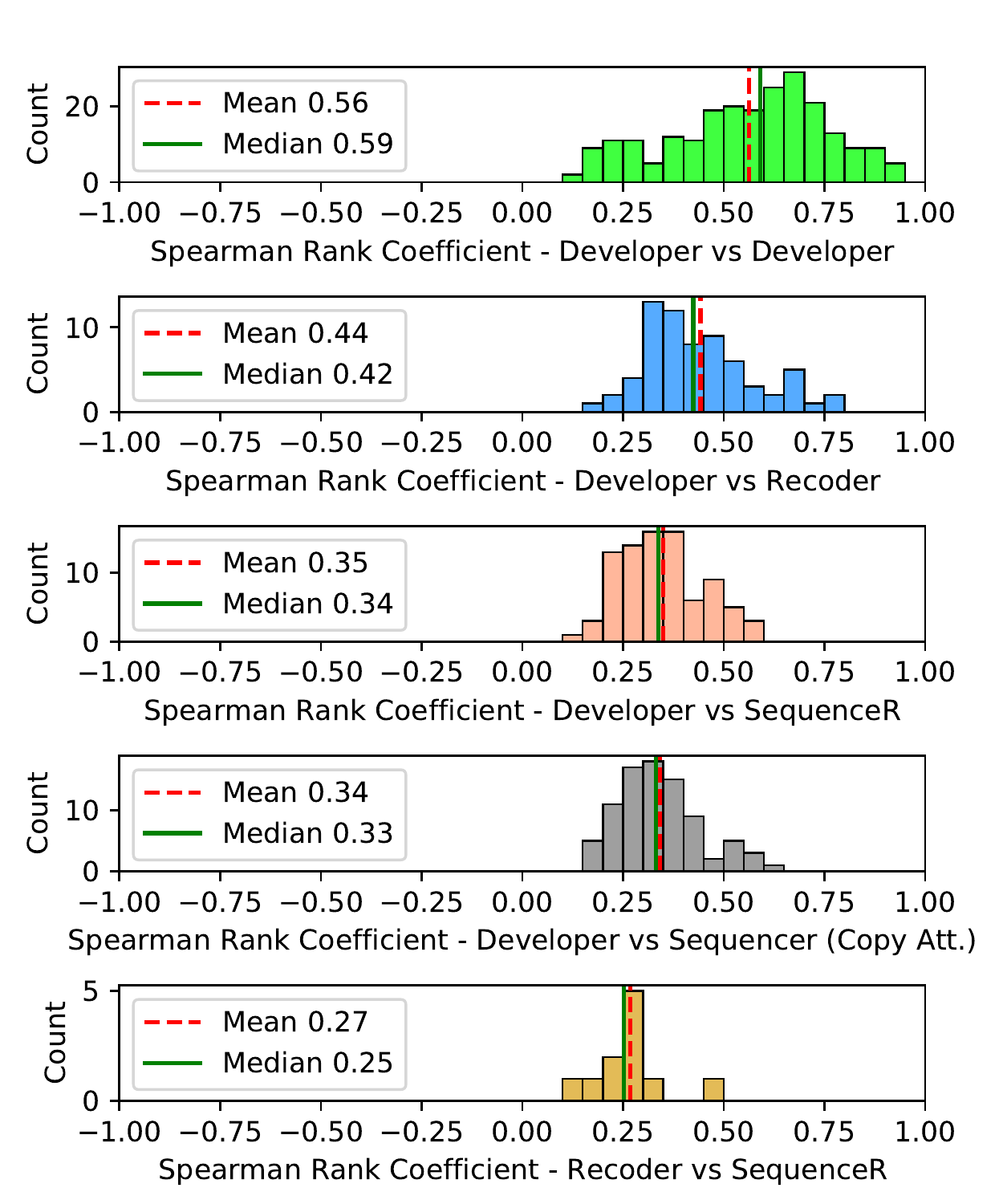}
  \caption{Correlation between attention paid by humans and models. $n$ is the number of pairs shows in the plot, i.e., after discarding statistically insignificant data points.}
  \label{fig:correlation_distribution}
\end{figure}

Figure~\ref{fig:correlation_distribution} shows the distribution of the correlation coefficients computed for different pairs of developers and models.
The top-most plot compares different developers with each other, which is a proxy for the inter-rater agreement.
The developer-developer correlation is generally high ($\mu_{dev} = \humanMeanAgreement{}$), indicating that different developers mostly agree on which tokens to focus on.

In the second and third plot of Figure~\ref{fig:correlation_distribution}, we show the distribution of correlation coefficients between developers and the two APR models.
The plot shows a positive correlation for both Recoder and SequenceR.
With a mean agreement of $\mu_{rec} = \recoderMeanAgreement{}$, Recoder is closer to the developers, whereas SequenceR reaches $\mu_{seq} = \sequencerMeanAgreement{}$.
As SequenceR computes also a copy attention vector, the fourth plot shows the correlation of copy attention with the developers.
We find regular attention and copy attention to be very similar for SequenceR (mean correlation between them: $0.89$), and hence focus on the regular attention of SequenceR in the remainder of this section.
Finally, the last plot of Figure~\ref{fig:correlation_distribution} gives the distribution of correlations between Recoder and SequenceR, which shows a positive but relatively low correlation of \modelvsmodelMeanAgreement{}.
As we consider only the attention from the most likely prediction of each model, there is a lower number of compared data points than in the previous histograms.

\begin{figure}[t]
  \centering
  \includegraphics[width=\linewidth]{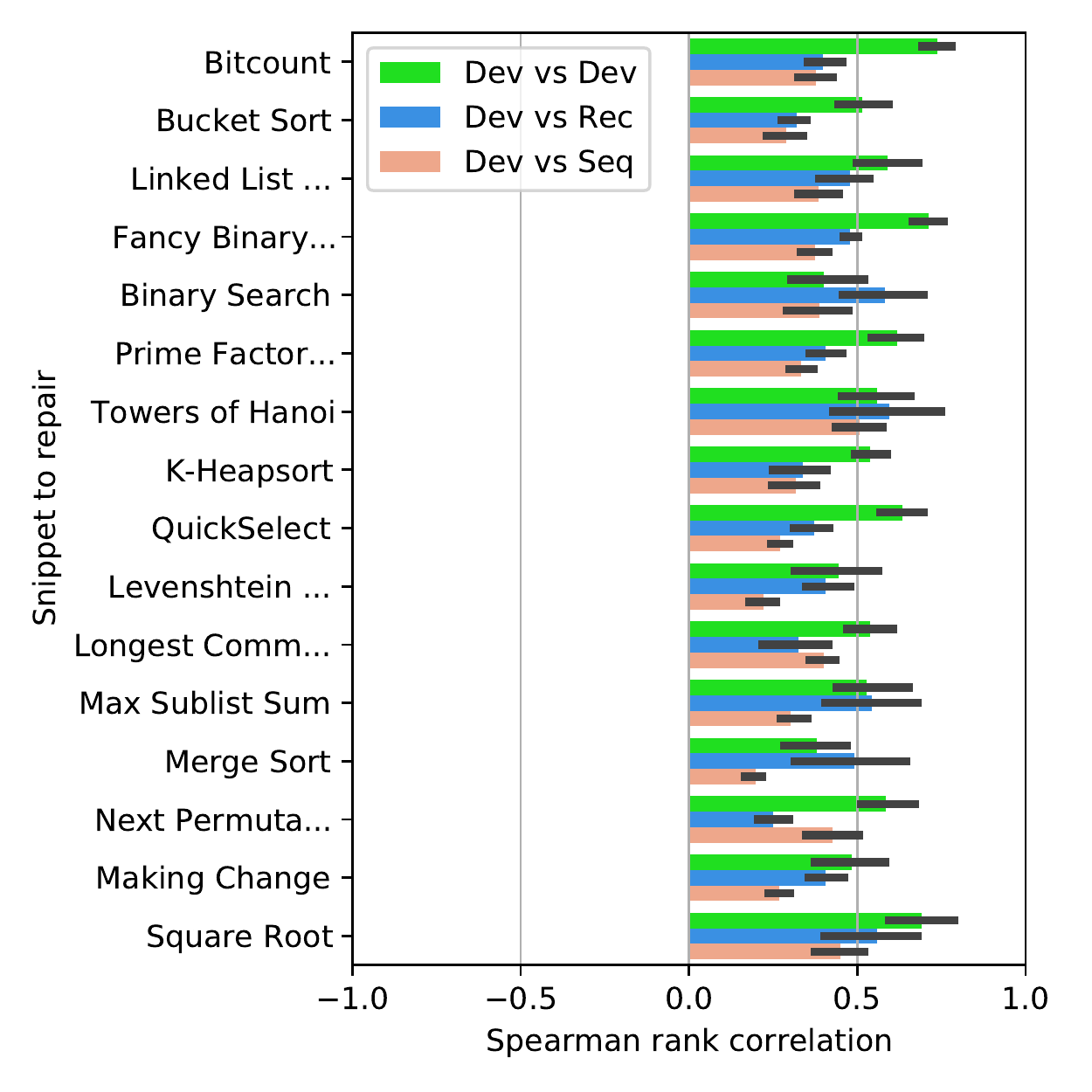}
  \caption{Correlation grouped per code snippet.}
  \label{fig:correlation_per_file}
\end{figure}

For a more fine-grained view on these results, Figure~\ref{fig:correlation_per_file} shows the correlation metrics for each individual \snippet{}.
The figure confirms the overall trends observed above.

\begin{answerbox}
  \textbf{Answer to RQ2}: Human developers mostly agree on what to look at when fixing a bug ($\mu_{dev} = \humanMeanAgreement{}$).
  The attention of the two neural APR models shows a relatively large agreement with the developers ($\mu_{rec} = \recoderMeanAgreement{}$ and $\mu_{seq} = \sequencerMeanAgreement{}$).
\end{answerbox}

\textit{Implications:}
Our findings have two implications.
First, higher attention agreement with human developers may indicate a more effective APR approach.
Recoder is shown to outperform SequenceR~\cite{zhuSyntaxguidedEditDecoder2021} and has a significantly higher mean correlation with humans than SequenceR (t-test with p-value: \pvalRecoderVsSequencerTestOnMeanAgreement{}).
A possible explanation is that Recoder's AST-based view on the code is similar to the information available to a developer.
For example, both developers and Recoder do not have to focus on curly brackets since the indentation already conveys this information to the developers, and the AST conveys it to Recoder.
In contrast, a sequence-based model, such as SequenceR, needs to infer such structural information from the given tokens.
Second, despite being positively correlated with the humans, both models do not yet reach the level of agreement that we see in the human-human correlation.
This finding suggests an opportunity for further improving models by more closely mimicking human attention.
Work on visual question answering~\cite{DBLP:journals/corr/abs-2109-13139} shows that using human-like attention while training a model achieves a new state of the art.
Future research could investigate a similar approach for neural APR models.

\subsection{RQ3: Focus on Buggy Line vs.\ Context}
The fact that many APR techniques assume the bug location to be given raises the question how much attention to pay to this location, as opposed to the surrounding code.
We investigate whether the buggy line and the surrounding context attract the same interest of developers and neural models.

\begin{figure}[t]
  \centering
  \includegraphics[width=\linewidth]{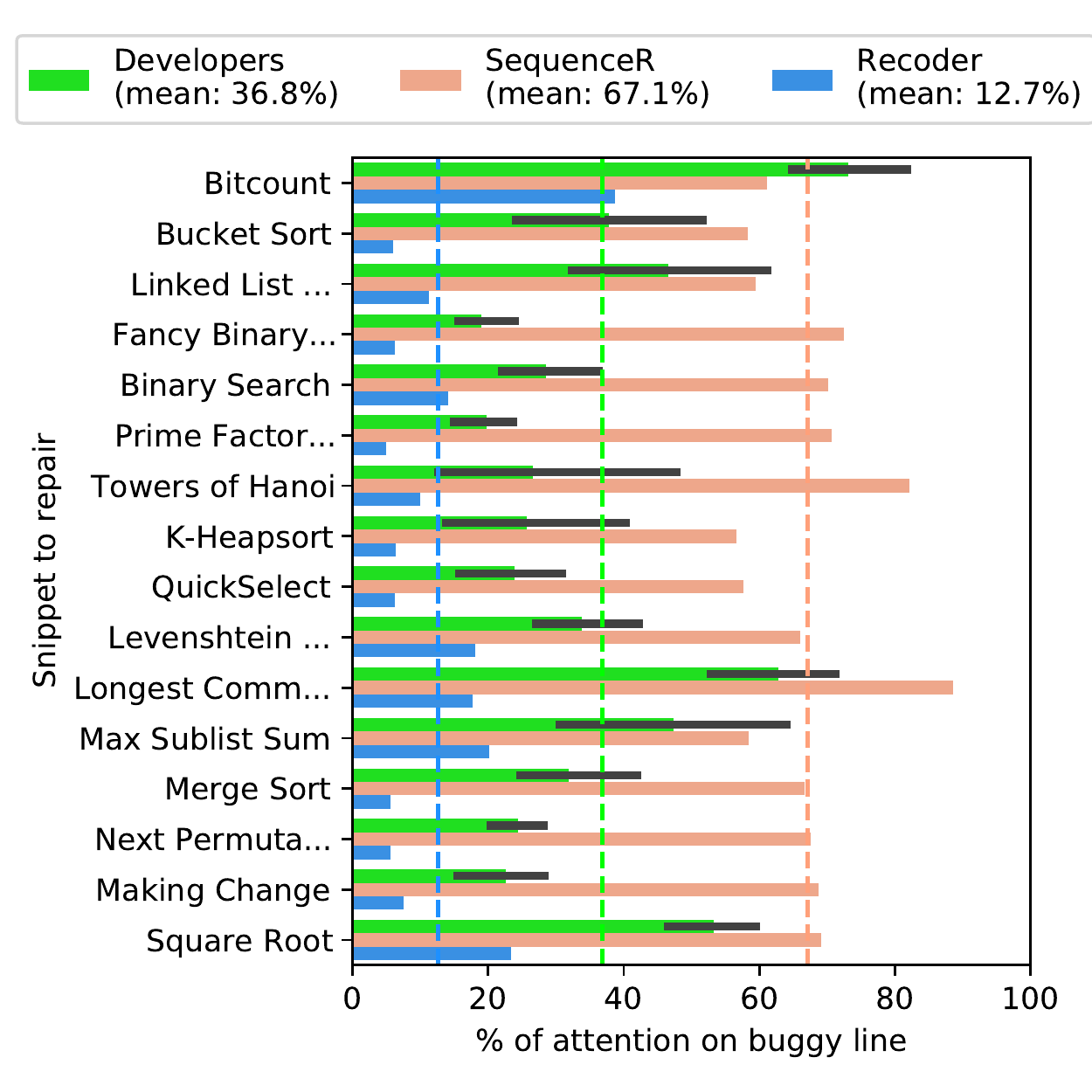}
  \caption{Attention given to the buggy line for developers and neural APR.}
  \label{fig:att_buggy_line}
\end{figure}

Figure~\ref{fig:att_buggy_line} shows the percentage of attention given to the buggy line by developers and neural APR models, grouped by \snippet{}.
The plot indirectly shows also the attention paid to the context, which is the attention left to reach 100\%.
SequenceR gives the highest amount of attention to the buggy line (mean: \sequencerAttentionBuggyLinePerc{}), whereas Recoder pays the least attention (mean: \recoderAttentionBuggyLinePerc{}), instead focusing mostly on the context.
The developers show a more balanced behavior exploring both buggy line (mean: \humanAttentionBuggyLinePerc{}) and context (mean: \humanAttentionContextPerc{}).

To further explore what drives the developers' attention to the buggy line or to the context, we study how the code length influences the percentage of attention given to the context.
Our hypothesis is that the context of a longer \snippet{} gets more attention than that of a short snippet.
To test this hypothesis, we compute the Spearman Rank correlation coefficient between the length of the \snippet{} in terms of number of tokens and the percentage of attention given to the context.
A statistically significant positive correlation of \corrContextVsCodeLength{} (p-value of \pvalContextVsCodeLength{}) between the two variables confirms the trend where developers explore the context of longer code snippets for a longer fraction of time than for shorter code snippets.

\begin{answerbox}
  \textbf{Answer to RQ3}: The two models give opposite answers to the question whether to pay more attention to the buggy line or to the surrounding context. While SequenceR spends the majority of its attention on the buggy line (\sequencerAttentionBuggyLinePerc{}), Recoder mostly focuses on the context (\recoderAttentionContextPerc{}).
  The developers instead show a hybrid behavior with \humanAttentionBuggyLinePerc{} of attention on the buggy line and \humanAttentionContextPerc{} on the surrounding context.
\end{answerbox}

\textit{Implications:}
Whether a model pays more or less attention to the buggy line is likely linked to its
neural model architecture.
A Seq2Seq model, as in SequenceR, predicts the entire new line of code to replace the buggy line with, which often is very similar except for a few tokens.
A possible explanation for SequenceR's high attention to the buggy line is that it needs to copy a large portion of tokens from the buggy line to the output.
This is not the case for Recoder, which is based on an AST representation of code and creates the patched line via tree transformations.
This observation motivates work on human-inspired neural models that adopt a more hybrid approach, similar to what we observe for developers.
Such a model could balance the attention on the buggy line and the context, e.g., as a function of the input code length.

\subsection{RQ4: Attention on Token Types}

Knowing which tokens are looked at by the developers and by the models, might reveal some blind spots of the current neural APR models.
To this end, we compute the \textit{distance from uniformity} (DFU) metric~\cite{paltenghiThinkingDeveloperComparing2021}, which measures how much attention a group of tokens is getting compared to a uniform distribution of attention.
A DFU $>$ 1 indicates that the tokens get more than uniform attention, DFU $<$ 1 means less, and DFU = 1 means the tokens get exactly uniform attention.
In other words, higher values mean more attention.
The DFU metric accounts for the fact that more common kinds of tokens get more absolute attention by normalizing the metric based on the number of occurrences of tokens.

\begin{figure}[t]
  \centering
  \includegraphics[width=\linewidth]{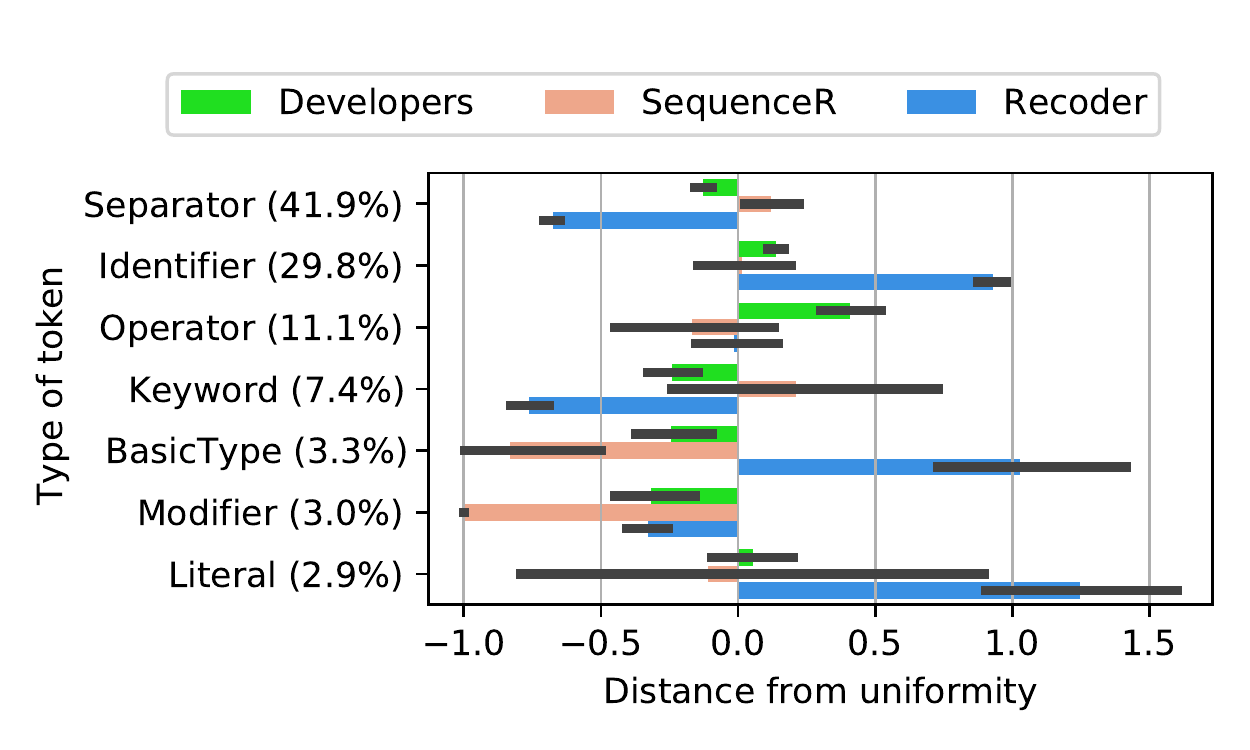}
  \caption{Distance from uniformity for token families. Higher values mean more attention.}
  \label{fig:dfu_token_families}
\end{figure}

Figure~\ref{fig:dfu_token_families} shows the DFU for different groups of tokens, each for the developers and the two studied models.
The label of each token group indicates the average proportion of code such tokens occupy across all the code snippets.
The figure shows that developers pay clearly more than uniform attention to identifiers and operators.
In contrast, neither SequencerR nor Recoder pay special attention to operators, signaling a possible blind spot of these models.
Recoder stands out by paying more than uniform attention to identifiers, types, and literals, perhaps in the hope they can carry a special meaning.
Some kinds of tokens, e.g., modifiers, are largely ignored by both developers and the two models.

\begin{answerbox}
  \textbf{Answer to RQ4}: The developers pay high attention to operators, signaling their pivotal role in the bug fixing process, but the two studied models do not focus on them very strongly.
\end{answerbox}

\textit{Implications:}
The difference in the attention given to operators by neural APR models and developers reveals a possible blind-spot of the models.
Future research could build on this hypothesis and propose new repair models trained to give an additional weight to tokens deemed most relevant by developers.

\subsection{RQ5: Bug Fixing Behavior Over Time}

Our final research question investigates how the focus of the developers changes over time while fixing a bug, both in terms of what they look at and when they edit the code.
To this end, we divide the time of each bug fixing session in $n$ time bins, each of which lasts for a fixed percentage of the total session time.
We set $n=20$, i.e., each bin describes the developer's behavior during 5\% of the total time spent on a bug.

\begin{figure}[t]
  \centering
  \includegraphics[width=\linewidth]{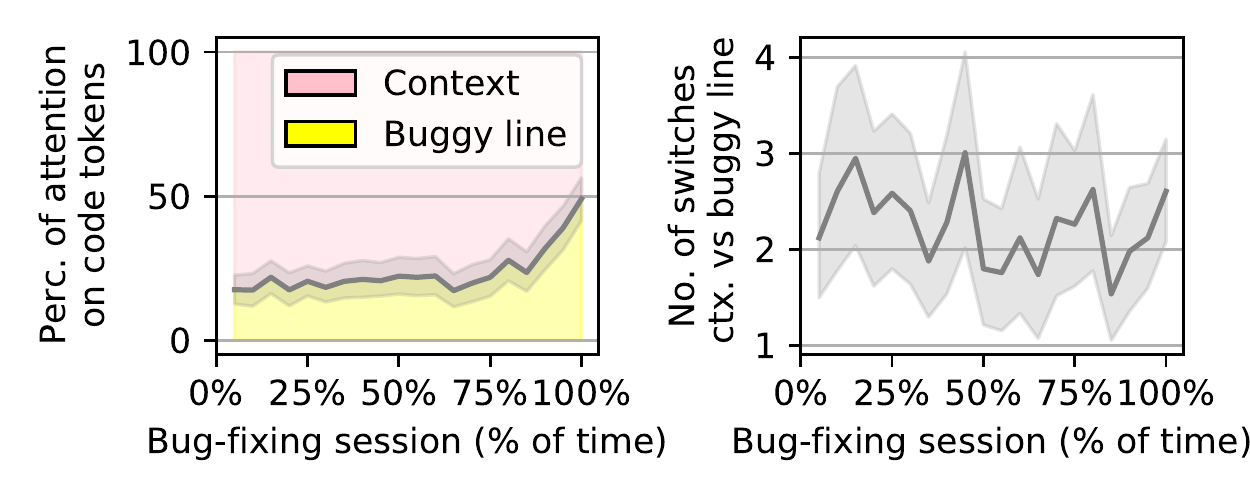}
  \caption{Attention given to context and buggy line over an average bug fixing session (on the left) and number of switches of attention between buggy line and context (on the right).}
  \label{fig:n_ctx_buggy_line_and_switches}
\end{figure}

In a first experiment, we study how the distribution of developer attention between the buggy line and the surrounding context evolves.
Figure~\ref{fig:n_ctx_buggy_line_and_switches} shows on the left how much attention is given to the context as opposed to the buggy line over time during an average bug fixing session.
We observe that the developers spend roughly one third of their time on the buggy line, with a burst of attention on the buggy line at the end of the bug fixing session.
Beside the sheer amount of time spent on the buggy line and context, we also report in the right plot of Figure~\ref{fig:n_ctx_buggy_line_and_switches} how often the developers switch between the two.
Switching can be seen as a proxy of the importance of connecting the information in both parts of the code.
The plot shows how the developers constantly interconnect the information of the buggy line with the surrounding context, potentially inspiring new neural APR that explicitly model a strong and constant link between the two areas.

\begin{figure}[t]
  \centering
  \includegraphics[width=\linewidth]{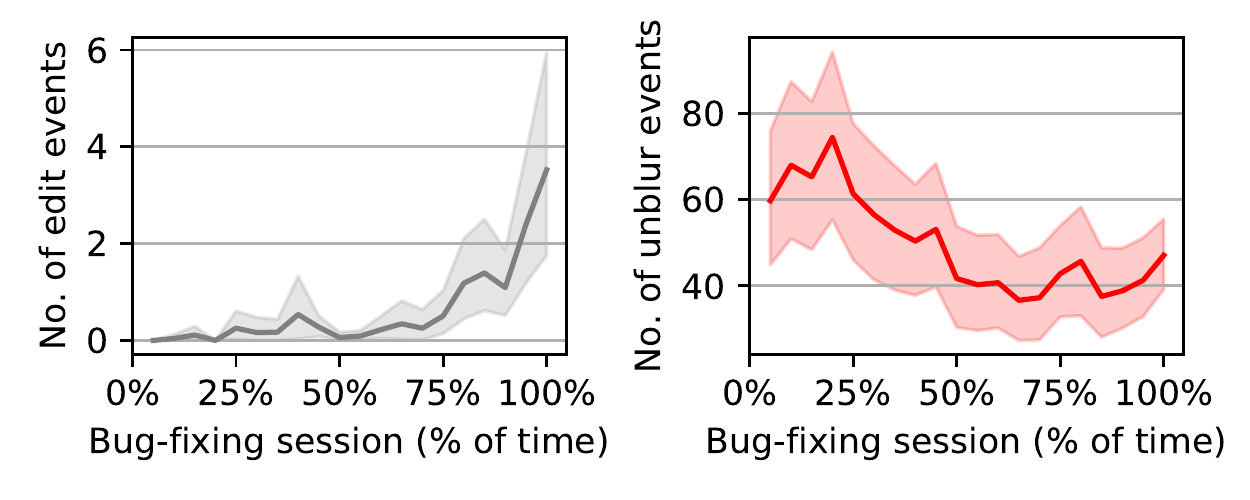}
  \caption{Number of edit and unblur events over an average bug fixing session.}
  \label{fig:n_edits_and_unblur_over_time}
\end{figure}

In a second experiment, we study how the amount of unblur and edit events evolves, which indicates whether developers explore or modify the code.
Figure~\ref{fig:n_edits_and_unblur_over_time}, left plot, shows the number of edit events, which remains relatively low for about 75\% of a bug fixing session and then clearly increases.
The plot on the right shows that the unblurring of code tokens decreases over time, indicating that developers are particularly active at inspecting the code during the first half of the bug fixing session.
Along with the left plot in Figure~\ref{fig:n_ctx_buggy_line_and_switches}, we interpret these results to indicate that the developers first explore code (i.e., many unblur events), then realize the problem, and after about 75\% of the total time start to fix the code (i.e., many edit events with high attention on the buggy line).

\begin{answerbox}
  \textbf{Answer to RQ5}: During an average bug fixing session, developers focus roughly one fourth of their attention on the buggy line and three fourth on the context, with constant switches between the two code areas.
  The first three quarters of the bug fixing session are devoted to code comprehension with a predominance of code inspection events, whereas the last fourth is dedicated to editing the code.
\end{answerbox}

\textit{Implications:}
Our findings lead to two implications.
First, the observed multi-phase behavior of developers motivates work on neural APR models that mimic how developers shift their attention over time.
Current APR models do not distinguish a code understanding and a code editing phase.
Second, the high number of switches between code in the buggy line and in the context motivates further research on neural APR models that explicitly model a direct and periodic connection between these two code parts.

\section{Threats to Validity}

\paragraph{Tracking Human Attention}
The following discusses factors that may impact the validity of our findings and, if possible, how we try to mitigate their effects.
One group of threats is about our way of tracking human attention.
First, the attention recorded via \interfaceNameAbbr{} may occasionally not accurately reflect what developers focus on, e.g., because a developer may move the cursor without looking at the screen.
We empirically validate this important part of our methodology in Section~\ref{sec:methodology_validation}.
Moreover, experiments that compare eye tracking with a similar unblurring-based interface show that unblurring can successfully approximate eye fixations~\cite{DBLP:journals/tochi/KimBBGODP17}, other work has used similar unblurring-based interfaces~\cite{paltenghiThinkingDeveloperComparing2021,muckeREyekerRemoteEye2021} for code-related tasks.
Second, the \interfaceNameAbbr{} interface might make the debugging process unnecessarily difficult.
The effectiveness results we obtain for developers hence may give an underestimate of the developers' abilities.
Third, participants may have cheated, e.g., by searching the web for a solution or by using an external IDE.
To minimize this risk, we rely exclusively on voluntary participants that we recruit through personal contacts.
Moreover, we manually check all attention maps to spot any obvious outliers.
\begin{figure}[t]
  \centering
  \includegraphics[width=\linewidth]{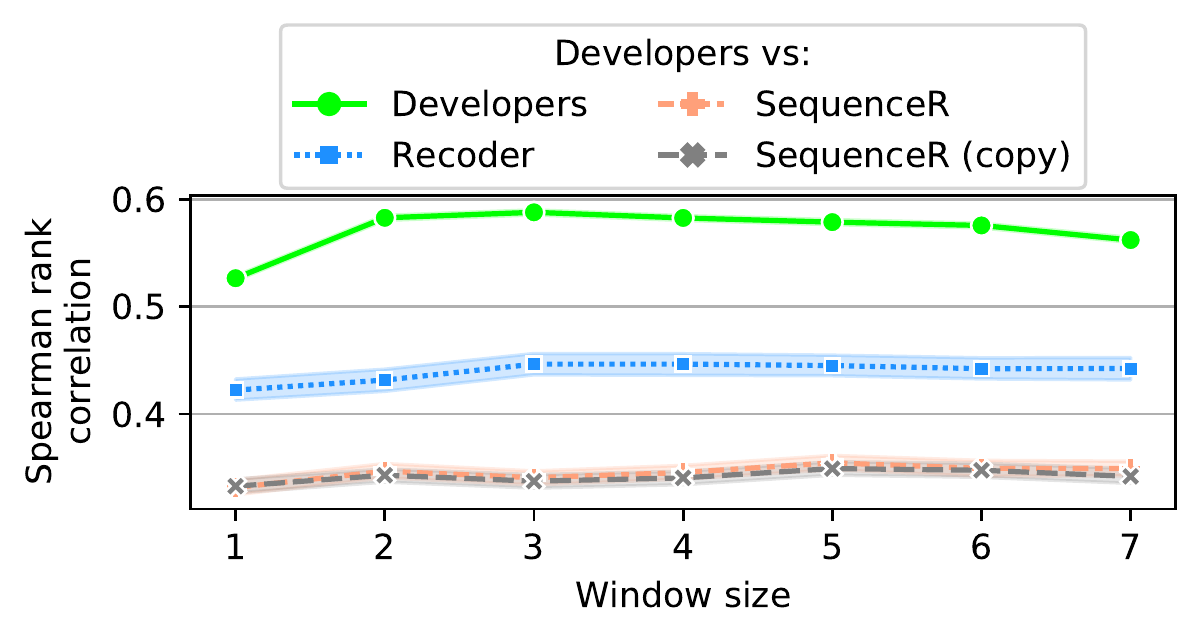}
  \caption{Effect of window size on the correlation score.}
  \label{fig:window_size_sensitivity}
\end{figure}
Finally, our methodology derives the attention on a window of seven contiguous tokens, but we cannot be sure on which of these tokens a developer actually focuses.
To understand the impact of the window size, we simulate the effect of window sizes smaller than seven by attributing attention only to a random subset of contiguous tokens among those in each window.
Figure~\ref{fig:window_size_sensitivity} shows how varying the window size between one and seven tokens affect some of the key results of our study (Section~\ref{sec:correlation results}).
We find that, except for a single-token window, the choice of the window size only marginally affects the results.

\paragraph{Attention of Neural Models}
Another threat concerns the attention of neural models, which has been criticized for providing noisy explanations about which parts of a model's input matter most for making a prediction~\cite{Jain2019,serranoAttentionInterpretable2019a}.
Others show that attention may not provide the one and only explanation for why a model makes a particular prediction, but it at least provides one plausible explanation~\cite{Wiegreffe2019}.
Despite not being perfect, attention is widely used to interpret the predictions of otherwise opaque models~\cite{wang2016attention,lee2017interactive,lin2017structured,ghaeini2018interpreting,ding2017visualizing}.
Moreover, \citet{rabinUnderstandingNeuralCode2021a} has recently shown how Sivand, an explainablility technique based on program simplification inspired by delta debugging~\cite{zellerSimplifyingIsolatingFailureinducing2002}, pinpoints important tokens that largely overlap with those reported by the attention mechanism.
Thus, due to its popularity and the recently shown connection with other explainability measures, we also rely on it here.

\paragraph{Other Threats}
Since this study considers relatively small Java programs representing stand-alone algorithms, the results of this work may not generalize to other programming languages or to more complex programs.
Finally, the studied APR techniques use test cases for validation, but the humans do not.
However, this difference does not impact our results, as validation is performed only after the developers and models propose a fix.

\section{Related Work}

\paragraph{Studies of code comprehension and debugging}
Castelhano et al.~\cite{castelhanoRoleInsulaIntuitive2019} study 20 subjects using eye tracking data while exploring buggy code snippets. They limit their analysis to the bug detection phase, whereas we focus on bug fixing.
Another study runs a remote experiment with twelve practitioners to study how they debug and fix bugs in C code~\cite{bohmeWhereBugHow2017a}.
Their work relies on surveys and self-reports from developers, without recording any of their coding sessions.
In contrast, besides involving a larger number of developers, our work records each and every change in the code editor during the entire bug fixing process.
Alaboudi et al.~\cite{alaboudiExploratoryStudyDebugging2021a} annotate recorded live-coding sessions from  developers and describe the alternation between different activities, such as debugging and programming.
Our findings in RQ5 partially confirms theirs based on fine-grained, automatically
tracked attention records.
Recent work~\cite{richterAreNeuralBug2023} compares developers to neural models on variable misuse bugs, but focuses on bug detection, whereas our work is about repair.

\paragraph{Studies of bug fixing and repair techniques}
Wang et al.~\cite{wangHowDifferentIt2019} study the patches generated by ten search-based and synthesis-based APR methods, and qualitatively compare the produced patches to those created by developers~\cite{justDefects4JDatabaseExisting2014}.
They focus on the final human patch, whereas we study the process of producing patches.
Moreover, we are the first to study the bug fixing process compared to neural APR models.
Another study~\cite{zhuMeaCulpaHow2021a} is about developer-produced patches for bugs in the SStuBs dataset~\cite{karampatsisHowOftenSingleStatement2020}.
They note that fixes done by an author of the buggy code tends to arrive faster and within a larger commit, compared to fixes created by other developers.
They do not observe the bug fixing process and, unlike us, have only one human-provided patch per bug.
\citet{zhangProgramRepairAutomated2022} study the human debugging process compared to three non-deep learning-based APR techniques for eight Defect4J bugs, whereas we study \nParticipants{} developers for \nBugFixes{} bugs and focus on learning-based techniques.

\paragraph{Developer and neural model attention}
Paltenghi et al.\ also compare human attention to neural models of code~\cite{paltenghiThinkingDeveloperComparing2021}.
However, they focus on code summarization, which does not require any human interactivity.
As a result, our interface for tracking human attention differs by providing not only a static image of the code, but an interactive code editor.
Due to the different tasks, we also consider different neural models, ask different research questions, and arrive at different results and implications.
Mucke et al.~\cite{muckeREyekerRemoteEye2021} propose a similar attention-tracking tool, but without any experimental evaluation.
Alternatives to tracking attention via unblurring include eye tracking~\cite{Rodeghero2015,Fakhoury2021}, fMRI-based experiments~\cite{DBLP:conf/icse/SiegmundKAPBLSB14}, and their combination~\cite{peitekSimultaneousMeasurementProgram2018}.
A key benefit of \interfaceNameAbbr{} is its lightweight setup, which facilitates remote participation.
Sharma et al.~\cite{sharmaExploratoryStudyCode2022} study how the self-attention of BERT~\cite{devlinBERTPretrainingDeep2019}, but they study only a pre-trained model, without focusing on a specific task.
\citet{paltenghiExtractingMeaningfulAttention2022} recently compare the attention of humans collected via eye tracking to that of pre-trained models, but the task is related to general code understanding, whereas we focus on bug fixing.

\paragraph{Automated program repair}
There is a wide range of APR approaches~\cite{cacm2019-program-repair}, using both traditional techniques and deep learning, including Recoder~\cite{zhuSyntaxguidedEditDecoder2021}, SequenceR~\cite{chenSequenceRSequencetoSequenceLearning2021},
CODIT\cite{chakrabortyCODITCodeEditing2020}, DLFix~\cite{liDLFixContextbasedCode2020b}, CoCoNuT~\cite{lutellierCoCoNuTCombiningContextaware2020}, TBar~\cite{liuTBarRevisitingTemplatebased2019}, jGenProg~\cite{martinezAstorExploringDesign2019},
SemFix~\cite{Nguyen2013b},
SimFix~\cite{jiangShapingProgramRepair2018},
RSRepair~\cite{qiAnalysisPatchPlausibility2015}, Nopol~\cite{xuanNopolAutomaticRepair2017}.
Recoder has been shown to outperform state of the art APR approaches on single-hunk bugs of Defects4J~\cite{justDefects4JDatabaseExisting2014}, and the
IntroClass~\cite{durieuxIntroClassJavaBenchmark2972016} and QuixBugs~\cite{linQuixBugsMultilingualProgram2017} datasets.
We select SequenceR and Recoder as the subjects of our study as they represent different kinds of neural models and both have recently achieved state of the art results.
AlphaRepair~\cite{xiaLessTrainingMore2022} is a recent APR approach that uses a pre-trained language model to generate patches, which also was subject to an empirical study~\citet{xiaPracticalProgramRepair2022}.
Comparing humans with large language models is left for future work.

\paragraph{Attention-based models of code}
Attention-based models solve various code-related tasks, e.g., code summarization~\cite{allamanisConvolutionalAttentionNetwork2016, ahmadTransformerbasedApproachSource2020a} and program repair~\cite{lutellierCoCoNuTCombiningContextaware2020,chenSequenceRSequencetoSequenceLearning2021,zhuSyntaxguidedEditDecoder2021}.
In parallel to the thrive of large language models for natural language, analogous models have been proposed in the programming language community as well, such as CodeBERT~\cite{fengCodeBERTPreTrainedModel2020}, Codex~\cite{chenEvaluatingLargeLanguage2021}, and AlphaCode~\cite{liCompetitionLevelCodeGeneration2022}.
Despite recent work~\cite{prennerCanOpenAICodex2022, sobaniaAnalysisAutomaticBug2023} applying Codex and ChatGPT to the QuixBugs dataset, we could not study its attention weights since the models are closed-source.

\section{Conclusion}
This paper studies two neural APR techniques and how their way of reasoning about the code compares to human developers.
By gathering a detailed record of \nDebuggingSessions{} bug fixing sessions via our novel \interfaceName{}, we perform a systematic study of the attention paid during the bug fixing process and how it affects the bug fixing effectiveness.
Our findings help better understand what current neural models do, do well, and perhaps should do in addition.
We find that the attention of the humans and both neural models often overlaps (\sequencerMeanAgreement{} to \recoderMeanAgreement{} correlation).
While the two models either focus mostly on the buggy line or on the surrounding context, the developers adopt a hybrid approach, where  \humanAttentionBuggyLinePerc{} of the attention is given to the buggy line and the rest to the context.
Overall, the humans still are clearly more effective at finding a correct fix (\humanAllCorrectPatchesPerc{} vs.\ less than 3\% correctly predicted patches).
Future work should investigate human-inspired APR methods, e.g., by mimicking the way humans distribute their attention to different parts of the code, and how humans split the bug fixing process into several phases.

\section{Data Availability}

The \interfaceNameAbbr{} programming interface,\footnote{\url{https://anonymous.4open.science/r/code-edit-recorder-arxiv}} our data analysis with complete dataset,\footnote{\url{https://anonymous.4open.science/r/study-developers-vs-apr/}} and our instrumented models\footnote{SequenceR: \url{https://figshare.com/s/52875aeb4f38cc3e21ca} and Recoder: \url{https://figshare.com/s/4b8c6ec59b2b85939b37}
} are publicly available and will be archived after publication.

\section*{Acknowledgements}
This work was supported by the European Research Council (ERC, grant agreement 851895), and by the German Research Foundation within the ConcSys and DeMoCo projects.

\bibliographystyle{ACM-Reference-Format}
\bibliography{phd-mattepalte,referencesMichael,references-dominik,moreRefs}

\end{document}